\documentclass[multphys,vecphys]{svmult}

\usepackage{makeidx}     
\usepackage{graphicx}    
\usepackage{multicol}    

\makeindex             

\begin{document}

\title*{{Observing the First Stars, One Star at a Time}}
\titlerunning{First GRBs}
\author{Abraham Loeb}
\institute{Astronomy Department, Harvard University, 60 Garden Street,
Cambridge, MA 02138;
\texttt{aloeb@cfa.harvard.edu}}
%
%
\maketitle

\section{Stellar Explosions at the Edge of the Visible Universe}
\label{sec:1}

Gamma-Ray Bursts (GRBs) are believed to originate in compact remnants
(black holes or neutron stars) of massive stars. Their high luminosities
make them detectable out to the edge of the visible universe
\cite{Reichart, Ciardi}.  GRBs offer the opportunity to detect the most
distant (and hence earliest\footnote{Observational cosmology resembles
archeology. By probing deeper into the universe, one reveals layers of it
that are more ancient (due to the finite speed of light).}) population of
massive stars, one star at a time.  In the hierarchical assembly process of
halos which are dominated by cold dark matter, the first galaxies should
have had lower masses (and lower stellar luminosities) than their
low-redshift counterparts.  Consequently, the characteristic luminosity of
galaxies or quasars is expected to decline with increasing redshift. GRB
afterglows, which already produce a peak flux comparable to that of quasars
or starburst galaxies at $z\sim 1-2$, are therefore expected to outshine
any competing source at the highest redshifts, when the first dwarf
galaxies have formed in the universe.

Preliminary polarization data from the {\it Wilkinson Microwave Background
Probe} ({\it WMAP}) indicates an optical depth to electron scattering of
$\sim 17\pm 4$\% after cosmological recombination \cite{WMAP}. This implies
that the first stars must have formed at a redshift $z\sim 20$ \cite{WL03,
Cen, Ciardi03, Somerville03} and reionized a substantial fraction of the
intergalactic hydrogen around that time. Early reionization can be achieved
with plausible star formation parameters in the standard $\Lambda$CDM
cosmology; in fact, the required optical depth can be achieved in a variety
of very different ionization histories (since {\it WMAP} places only an
integral constraint on these histories \cite{holder}). One would like to
probe the full history of reionization in order to disentangle the
properties and formation history of the stars that are responsible for
it. GRB afterglows offer the opportunity to detect stars as well as to
probe the ionization state \cite{Barkana03} and metal enrichment level
\cite{FL03} of the intervening intergalactic medium (IGM).

GRBs, the electromagnetically-brightest explosions in the universe, should
be detectable out to redshifts $z>10$ \cite{Reichart,Ciardi}.
High-redshift GRBs can be easily identified through infrared photometry,
based on the Ly$\alpha$ break induced by absorption of their spectrum at
wavelengths below $1.216\, \mu {\rm m}\, [(1+z)/10]$. Follow-up
spectroscopy of high-redshift candidates can then be performed on a
10-meter-class telescope. There are four main advantages of GRBs relative
to traditional cosmic sources such as quasars:

\begin{itemize}

\item The GRB afterglow flux at a given observed time lag after the
$\gamma$-ray trigger is not expected to fade significantly with
increasing redshift, since higher redshifts translate to earlier times
in the source frame, during which the afterglow is intrinsically
brighter \cite{Ciardi}. For standard afterglow lightcurves and
spectra, the increase in the luminosity distance with redshift is
compensated by this {\it cosmological time-stretching} effect.

\item As already mentioned, in the standard $\Lambda$CDM cosmology,
galaxies form hierarchically, starting from small masses and increasing
their average mass with cosmic time. Hence, the characteristic mass of
quasar black holes and the total stellar mass of a galaxy were smaller at
higher redshifts, making these sources intrinsically fainter
\cite{Wyithe}. However, GRBs are believed to originate from a stellar mass
progenitor and so the intrinsic luminosity of their engine should not
depend on the mass of their host galaxy. GRB afterglows are therefore
expected to outshine their host galaxies by a factor that gets larger with
increasing redshift.

\item Since the progenitors of GRBs are believed to be stellar, they likely
originate in the most common star-forming galaxies at a given redshift
rather than in the most massive host galaxies, as is the case for bright
quasars \cite{Barkana03}. Low-mass host galaxies induce only a weak
ionization effect on the surrounding IGM and do not greatly perturb the
Hubble flow around them. Hence, the Ly$\alpha$ damping wing should be
closer to the idealized unperturbed IGM case \cite{jordi98} and its
detailed spectral shape should be easier to interpret. Note also that
unlike the case of a quasar, a GRB afterglow can itself ionize at most
$\sim 4\times 10^4 E_{51} M_\odot$ of hydrogen if its UV energy is $E_{51}$
in units of $10^{51}$ ergs (based on the available number of ionizing
photons), and so it should have a negligible cosmic effect on the
surrounding IGM.

\item GRB afterglows have smooth (broken power-law) continuum spectra
unlike quasars which show strong spectral features (such as broad emission
lines or the so-called ``blue bump'') that complicate the extraction of IGM
absorption features. In particular, the continuum extrapolation into the
Ly$\alpha$ damping wing (the Gunn-Peterson \cite{GP} absorption trough)
during the epoch of reionization is much more straightforward for the
smooth UV spectra of GRB afterglows than for quasars with an underlying
broad Ly$\alpha$ emission line \cite{Nat,Barkana03}.

\end{itemize}

Although the nature of the central engine that powers the relativistic jets
of GRBs is unknown, recent evidence indicates that GRBs trace the formation
of massive stars \cite{bloom,kul,Tot97,Wij98,BlNat00}, and in particular
that long-duration GRBs are associated with Type Ib,c supernovae
\cite{Stanek}.  Since the first stars in the universe are predicted to be
predominantly massive \cite{Abel,BCL2002}, their death might give rise to
large numbers of GRBs at high redshifts.

\section{GRB Afterglows at High Redshifts}

\subsection{Motivation}

In difference from quasars of comparable brightness, GRB afterglows are
short-lived and release $\sim 10$ orders of magnitude less energy into the
surrounding IGM. Beyond the scale of their host galaxy, they have a
negligible effect on their cosmological environment\footnote{Note, however,
that feedback from a single GRB or supernova on the gas confined within
early dwarf galaxies could be dramatic, since the binding energy of most
galaxies at $z>10$ is lower than $10^{51}~{\rm ergs}$
\cite{review}.}. Consequently, they make ideal probes of the IGM during the
reionization epoch.  Their rest-frame UV spectra can be used to probe the
ionization state of the IGM through the spectral shape of the Gunn-Peterson
\cite{GP} (Ly$\alpha$) absorption trough, or its metal enrichment history
through the intersection of enriched bubbles of supernova ejecta from early
galaxies\cite{FL03}.  Afterglows that are unusually bright ($>10$mJy) at
radio frequencies should also show a detectable forest of 21 cm absorption
lines due to enhanced HI column densities in sheets, filaments, and
collapsed mini-halos within the IGM \cite{Carilli,FurLoeb}.

Another advantage of GRB afterglows is that once they fade away, one may
search for their host galaxies. Hence, GRBs may serve as signposts of the
earliest dwarf galaxies that are otherwise too faint or rare on their own
for a dedicated search to find them. Detection of metal absorption
lines from the host galaxy in the afterglow spectrum, offers an unusual
opportunity to study the physical conditions (temperature, metallicity,
ionization state, and kinematics) in the interstellar medium of these
high-redshift galaxies.

\begin{figure}
\centering
\includegraphics[height=10cm]{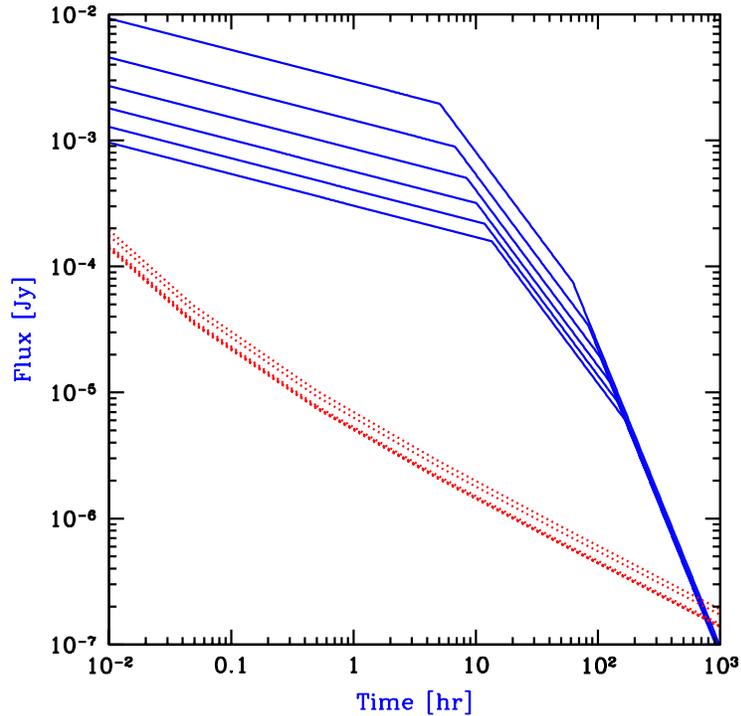} 
\caption{GRB afterglow flux as a function of time since the $\gamma$-ray
trigger in the observer frame (taken from \cite{Barkana03}). The flux
(solid curves) is calculated at the redshifted Ly$\alpha$ wavelength. The
dotted curves show the planned detection threshold for the {\it James Webb
Space Telescope} ({\it JWST}), assuming a spectral resolution $R=5000$
with the near infrared spectrometer, a signal to noise ratio of 5 per
spectral resolution element, and an exposure time equal to $20\%$ of the
time since the GRB explosion (see  http://www.ngst.stsci.edu/nms/main/~).
Each set of curves shows a sequence of redshifts, namely $z=5$, 7, 9, 11,
13, and 15, respectively, from top to bottom.}
\label{fig1}
\end{figure}

\subsection{Afterglow Flux at the Ly$\alpha$ Wavelength}

\begin{figure}
\centering
\includegraphics[height=10cm]{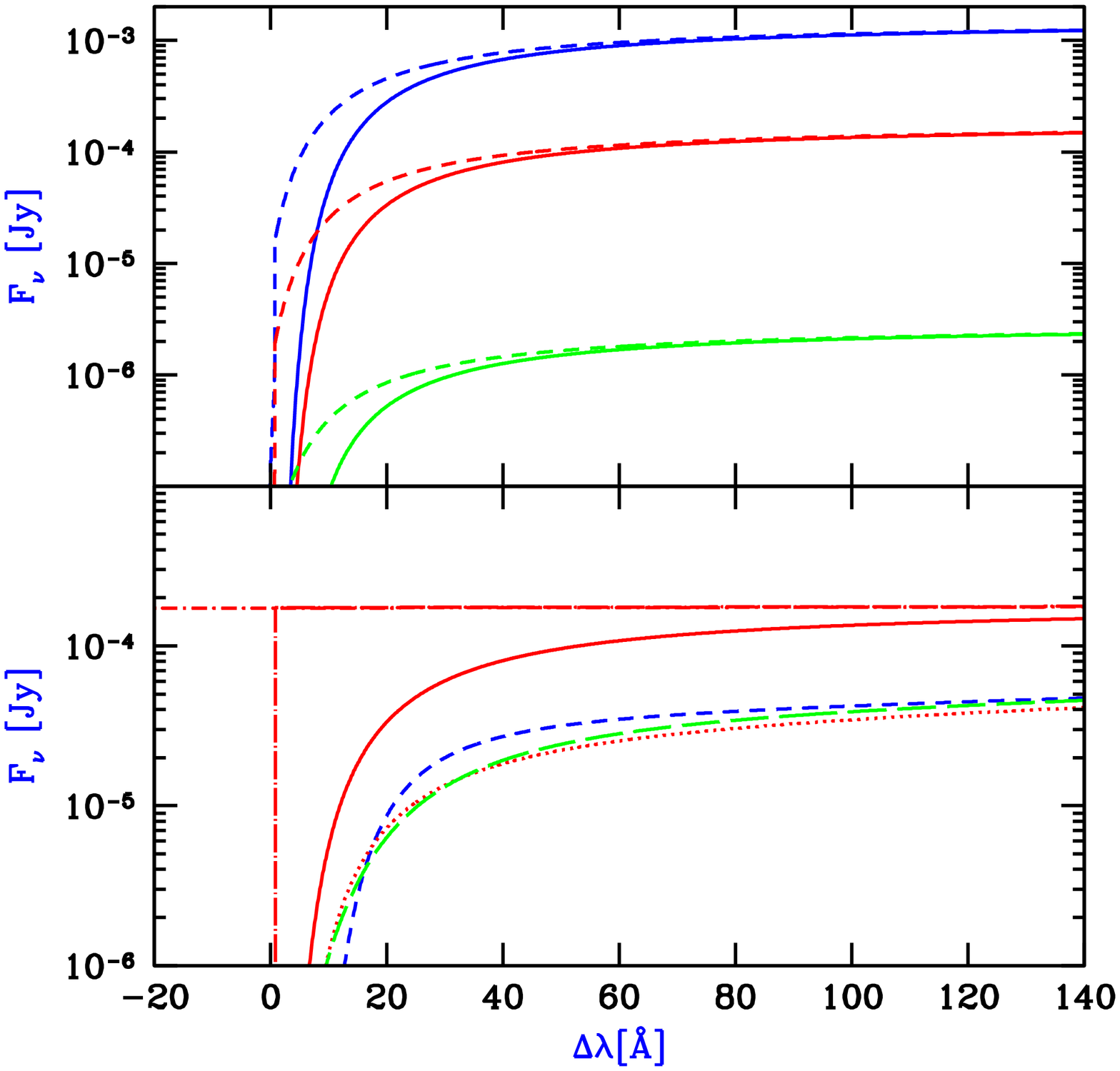} 
\caption{Expected spectral shape of the Ly$\alpha$ absorption trough due to
intergalactic absorption in GRB afterglows (taken from
\cite{Barkana03}). The spectrum is presented in terms of the flux density
$F_{\nu}$ versus relative observed wavelength $\Delta \lambda$, for a
source redshift $z=7$ (assumed to be prior to the final reionization phase)
and the typical halo mass $M=4 \times 10^8 M_{\odot}$ expected for GRB host
galaxies that cool via atomic transitions. {\it Top panel:} Two examples
for the predicted spectrum including IGM HI absorption (both resonant and
damping wing), for host galaxies with (i) an age $t_S=10^7$ yr, a UV escape
fraction $f_{\rm esc}=10\%$ and a Scalo initial mass function (IMF) in
solid curves, or (ii) $t_S=10^8$ yr, $f_{\rm esc}=90\%$ and massive
($>100M_\odot$) Pop III stars in dashed curves. The observed time after the
$\gamma$-ray trigger is one hour, one day, and ten days, from top to
bottom, respectively. {\it Bottom panel:} Predicted spectra one day after a
GRB for a host galaxy with $t_S=10^7$ yr, $f_{\rm esc}=10\%$ and a Scalo
IMF. Shown is the unabsorbed GRB afterglow (dot-short dashed curve), the
afterglow with resonant IGM absorption only (dot-long dashed curve), and
the afterglow with full (resonant and damping wing) IGM absorption (solid
curve). Also shown, with 1.7 magnitudes of extinction, are the afterglow
with full IGM absorption (dotted curve), and attempts to reproduce this
profile with a damped Ly$\alpha$ absorption system in the host galaxy
(dashed curves).  (Note, however, that damped absorption of this type could
be suppressed by the ionizing effect of the afterglow UV radiation on the
surrounding interstellar medium of its host galaxy\cite{PernaLoeb}.) Most
importantly, the overall spectral shape of the Ly$\alpha$ trough carries
precious information about the neutral fraction of the IGM at the source
redshift; averaging over an ensemble of sources with similar redshifts can
reduce ambiguities in the interpretation of each case due to particular
local effects.}
\label{fig2}
\end{figure}

Figure 1 shows the expected spectral flux from GRB afterglows as a function
of observed time after the GRB trigger for a sequence of redshifts
(assuming typical model parameters \cite{Barkana03}). The flux is
calculated at the rest-frame Ly$\alpha$ wavelength, where intergalactic HI
should produce the Gunn-Peterson \cite{GP} trough prior to reionization.
As already mentioned, the flux does not decline dramatically with
increasing redshift since cosmic time stretching counteracts the luminosity
distance increase.

The expected spectral shape of the Gunn-Peterson trough prior to the final
overlap phase of HII regions in the IGM, is shown in Figure 2.  If GRBs
trace the typical sites of star formation, then most of them should be
hosted by dwarf galaxies at high redshifts. Low-mass hosts would perturb
only weakly the surrounding IGM (radiatively through their ionizing flux,
gravitationally [through cosmological infall, or hydrodynamically through
their winds), in difference from the massive host galaxies of the brightest
quasars at the same epoch \cite{Nat, Barkana03}. Consequently, the spectral
shape of the Ly$\alpha$ trough (which is proportional to
$e^{-\tau(\lambda)}$, where $\tau(\lambda)$ is the optical depth as a
function of wavelength) is simpler to interpret for GRB afterglows than it
is for quasars. The optical depth as a function of wavelength in the
Ly$\alpha$ damping wing is linearly proprtional to the mean neutral
fraction of the IGM, $x_{\rm HI}$, since its amplitude is normalized by the
optical depth at the Ly$\alpha$ resonance, $\sim 6.75\times 10^5 x_{\rm
HI}[(1+z)/10]^{3/2}$ \cite{jordi98,review}.

\begin{figure}
\centering
\includegraphics[height=10cm]{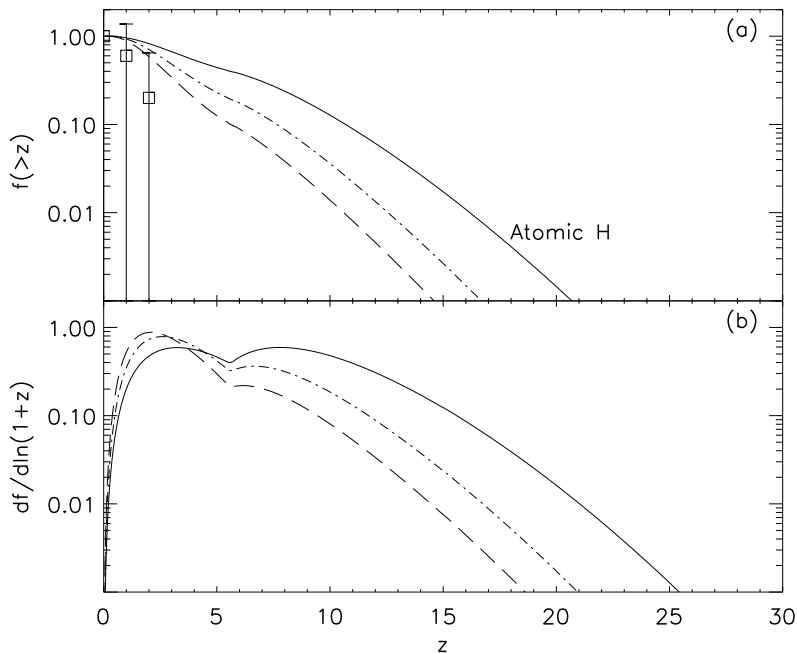}
\caption{Redshift distribution of all GRBs as compared to that measured by
flux-limited surveys (taken from \cite{BrL2002}).  ({\it a}) Fraction of
bursts that originate at a redshift higher than $z$ vs. $z$. The data
points reflect $\sim 20$ observed redshifts.  ({\it b}) Fraction of bursts
per logarithmic interval of $(1+z)$ vs. $z$.  {\it Solid lines}: All GRBs
for star formation through atomic line cooling.  {\it Dot-dashed lines}:
Expected distribution for {\it Swift}.  {\it Long-dashed lines}: Expected
distribution for {\it BATSE}.  Note that the curves for the two
flux-limited surveys are very uncertain due to the poorly-determined GRB
luminosity function. Nevertheless, GRB number counts could provide a rough
measure of the cosmic formation rate of massive stars at high redshifts. }
\label{fig3}
\end{figure}

\subsection{Statistics}

The upcoming {\it Swift} satellite (see http://swift.gsfc.nasa.gov/),
planned for launch by the end of 2003, is expected to detect about a
hundred GRBs per year. Figure 3 shows the expected redshift distribution of
all GRBs (solid lines), under the assumption that the GRB rate is simply
proportional to the star formation rate \cite{BrL2002}.  The long-dashed
and the dot-dashed lines show the same rate, but taking into account the
limited sensitivity of the {\it BATSE} and {\it Swift} detectors,
respectively. This calculation implies that about a quarter of all GRBs
detected by {\it Swift} should originate at $z > 5$. This estimate is
rather uncertain because of the poorly determined GRB luminosity function
\cite{BrL2002}.

\begin{figure}
\centering
\includegraphics[height=10cm]{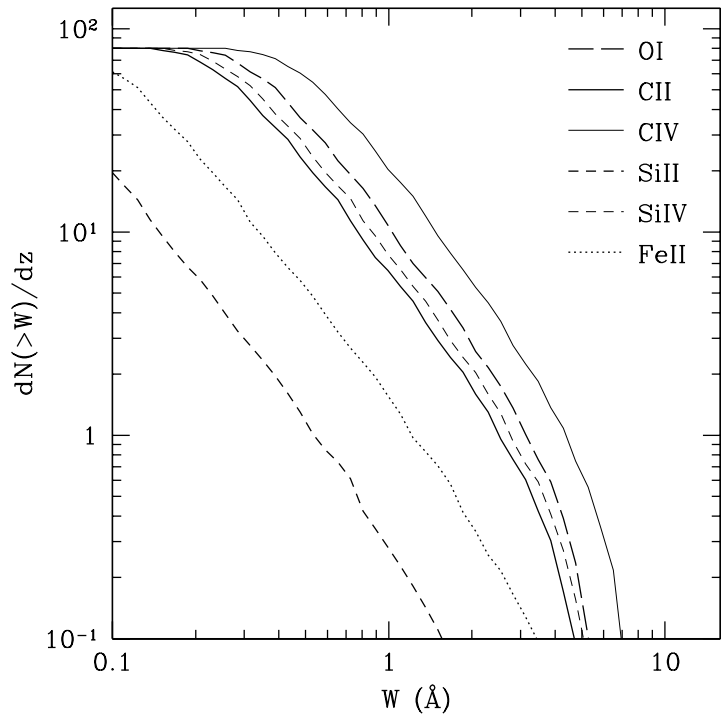}
\caption{Predicted number of line-of-sight intersections of metal-enriched
bubbles around the first galaxies per redshift interval at a redshift $z=8$
(taken from \cite{FL03}). This statistic was calculated as a function of
the equivalent width threshold of the metal absorption lines produced by
these bubbles, for typical parameter choices of the theoretical model (see
\cite{FL03} for details).  Thick solid, long-dashed, short-dashed, and
dotted curves are for C II, O I, Si II, and Fe II metal lines,
respectively.  Thin solid and short-dashed curves are for C IV and Si IV,
respectively.}
\label{fig4}
\end{figure}

In principle, the rate of high-redshift GRBs may be significantly
suppressed if the early massive stars fail to launch a relativistic
outflow. This is possible, since metal-free stars may experience negligible
mass loss before exploding as a supernova. They would then retain their
massive hydrogen envelope, and any relativistic jet might be quenched
before escaping the star. However, localized metal enrichment is expected
to occur rapidly (on a timescale much shorter than the age of the
then-young universe) due to starbursts in the first galaxies and so even
the second generation of star formation could occur in an interstellar
medium with a significant metal content, resulting in massive stars that
resemble more closely the counterparts of low-redshift GRB progenitors.

We note that the known population of optically-dark GRBs \cite{kul} is most
likely associated with dust-obscured GRBs at low redshifts, although a
small fraction of these bursts could be optically-dark due to Ly$\alpha$
absorption by the IGM at high-redshift.

\subsection{Progenitors}

In the previous section, we have assumed for simplicity a constant
efficiency of forming GRBs per unit mass of stars. This simplifying
assumption may either overestimate or understimate the frequency of
GRBs. Metal-free stars are thought to be massive \cite{Abel,BCL2002} and
their extended envelopes may suppress central jets within them (which may
be produced through the collapse of their core to a spinning black
hole). On the other hand, low-metallicity stars are expected to have weak
winds with little angular momentum loss during their evolution, and so they
may preferentially yield rotating central configurations that produce GRB
jets after core collapse \cite{Heg03,Fynbo}. The basic question: {\it
should GRBs be common in the first generation of metal-free stars?}
remains open for further study.

\section{{\it Swift}: A Challenge for Infrared Follow-Ups}

A small fraction ($\sim 10$--$25\%$) of the GRB afterglows detected by {\it
Swift} are expected to originate at redshifts $z>5$.  This subset of
afterglows can be selected photometrically using a small telescope, based
on the Ly$\alpha$ break at a wavelength of $1.216\, \mu {\rm m}\,
[(1+z)/10]$, caused by intergalactic HI absorption.  The challenge in the
upcoming years would be to follow-up on these candidates spectroscopically,
using a large (10-meter class) telescope.  A high-resolution spectrum can
then be used to trace the ionization state (see Fig. 2) and metal
enrichment state of the gas along the line-of-sight.  For example, Figure 4
illustrates the expected number of intersections of enriched bubbles of
supernova ejecta from early galaxies, as a function of the equivalent width
of the metal absorption lines they produce (see \cite{FL03} for
details). Measurement of this statistic for different metals can be used to
constrain the initial mass function (through element abundances) and the
mechanical energy output (through bubble sizes) of the first stars, as well
as the significance of molecular H$_2$ cooling in forming stars within
the first mini-halos (through the number density of small bubbles).

Based on the arguments mentioned in \S 1, GRB afterglows are likely to
revolutionize observational cosmology and replace traditional sources like
quasars, as probes of the IGM at $z>5$. We caution, however, that this is a
theoretical expectation (the highest redshift GRB detected so far
\cite{Andersen} is at $z\sim 4.5$) and as such it is subject to
uncertainties about the production of relativistic outflows by the first
generation of stars.

However, note that even if GRB outflows are quenched in the massive
envelopes of low-metallicity stars, one would expect them to appear at high
redshifts. This follows from the fact that metallicity obtains a large
range of possible values (in different spatial regions) at any slice of
cosmic time.  For example, the broad emission lines of all known quasars
show evidence for high metallicity gas \cite{Hamann,Dietrich}, and so it is
clear that there are pockets of highly enriched gas near the centers of
massive galaxies at $z>6$, where star formation reached an advanced stage
and GRB production should be as likely as it is in the local universe.  The
cores of the most massive galaxies at $z>6$ simply predate the
characteristic evolution of the rest of the universe at a much lower
redshift, because these cores reside in unusually overdense cosmological
environments \cite{LoebPeebles}.  Such environments are likely to produce
GRB progenitors in the same way that less overdense cosmological regions
produce them at a lower redshift. {\it Swift} has the potential to educate
us whether GRBs are produced only within these massive galaxies or in the
more common sites of star formation at early cosmic times, namely the
low-metallicity dwarf galaxies.

In either case, the near future promises to be exciting for GRB astronomy
as well as for studies of the high-redshift universe.

\bigskip
\bigskip
\bigskip
\noindent {\bf Acknowledgements}

I am indebted to my collaborators for the various calculations mentioned in
this overview: Rennan Barkana, Volker Bromm, Benedetta Ciardi, Steve
Furlanetto, Jim Peebles, and Stuart Wyithe. This work was supported in part
by NSF grants AST-0204514, AST-0071019, NASA grant ATP02-0004-0093, NATO
grant PST.CLG.979414, and the John Simon Guggenheim Memorial Fellowship.

%
%

%
%

\printindex
\end{document}